# High-$T_c$ Superconductors - based Nanocomposites with Improved Intergrain Coupling and Enhanced Bulk Pinning

Tanya Prozorov, Brett McCarty, Baowei Liu and Ruslan Prozorov

*Abstract* — Heterogeneous sonochemical synthesis was used to modify superconducting properties of granular $YBa_2Ca_3CuO_{7-\delta}$ and $Bi_2Sr_2CaCu_2O_{8+x}$. Sonication of liquid-powder alkane slurries produces material with enhanced intergrain coupling and improved current-carrying capabilities. Co-sonication with metals and organometallics results in highly compact nanocomposites with increased magnetic irreversibility. Ultrasonic irradiation of $YBa_2Ca_3CuO_{7-\delta}$ carried under partial oxygen atmosphere produces similar morphological effects and *increases* superconducting transition temperature due to effective surface saturation with oxygen. Detailed chemical and physical characterization of sonochemically prepared high-$T_c$ nanocomposites is presented.

## I. Introduction

USEFUL properties of superconductors, such as critical current, critical fields and magnetic irreversibility, strongly depend on the material's morphology. High-$T_c$ cuprates, $YBa_2Ca_3CuO_7$ (YBCO) and $Bi_2Sr_2CaCu_2O_8$ (BSCCO) are potentially useful mostly in their bulk form. However, achieving high persistent currents in these granular materials is a non-trivial task [1]. BSCCO is highly anisotropic, and very hard and brittle, and critical currents are limited mostly by the intergrain coupling. YBCO, on the other hand, is very sensitive to the oxygen content and distribution, which limits possible technological treatments.

This paper describes a novel method for modification of microstructure and introduction of efficient pinning centers in high-$T_c$ superconductors. The presented method utilizes high-intensity ultrasonic irradiation for structure modification and preparation of nanocomposites based on high-$T_c$ superconductors.

## II. Experimental

### A. Sonochemical Method

Irradiation of liquids with powerful ultrasound induces transient cavitation: nucleation, growth and violent collapse of bubbles [2,3]. The implosive bubble collapse generates localized hot spots with temperatures as high as 5000 K, pressures of about 800 atm, and cooling rates exceeding $10^9$ K/s, and induces intense shock waves, propagating in the liquid at velocities well above the speed of sound. In powder-liquid mixtures (slurries), shockwaves lead to an extremely rapid mass transfer and induce high velocity collisions among suspended solid particles [2-4]. Such interparticle collisions result in extreme heating at the point of impact, which can lead to effective localized melting and significant increase in the rates of numerous solid-liquid reactions. As a consequence, morphology of the initial material is significantly modified: individual grains are grinded, smoothened and welded together, ultimately resulting to a more compact material. In the case of a superconductor, such morphology change leads to better inter-grain coupling and annealing of the intra-grain defects. Sonication with volatile organometallic compounds leads to *in-situ* nucleation of nanoparticles, which precipitate on the surface of individual granules, and become trapped between colliding grains. The process is so aggressive that it overcomes usual surface tension limitations, and yields a uniform composite with nanoparticles embedded in the bulk of the slurry. This approach was initially explored in polycrystalline $MgB_2$. Resulted $MgB_2$-based nanocomposites with magnetic and non-magnetic embedded nanoparticles exhibited enhanced pinning. Both methods lead to production of novel nanocomposite superconducting materials [4]. Applied to polycrystalline BSCCO, irradiation with high-intensity ultrasound was shown to significantly improve current-carrying characteristics and magnetic irreversibility [5].

### B. Sample Preparation

Slurry of 2 wt% of polycrystalline $Bi_2Sr_2CaCu_2O_{8+x}$ (325

Manuscript received October 4, 2004. This work was supported in part by the donors of the American Chemical Society Petroleum Research Fund and USC Research & Productive Scholarship Award. The SEM study was carried out in the Center for Microanalysis of Materials (UIUC), which is partially supported by the DOE under Grant DEFGO2-91-ER45439.

T. Prozorov was with the School of Chemical Sciences, University of Illinois, Urbana, IL 61801. She is now with the Department of Chemical Engineering, University of South Carolina, Columbia, SC 29208 USA (e-mail: prozorot@engr.sc.edu).

B. McCarty is with the Department of Physics & Astronomy, University of South Carolina, Columbia, SC 29208 USA (e-mail: saiatusc@hotmail.com).

R. Prozorov is with the Department of Physics & Astronomy, University of South Carolina, Columbia, SC 29208 USA (corresponding author, phone: 803-777-8197; fax: 803-777-3065; e-mail: prozorov@sc.edu).

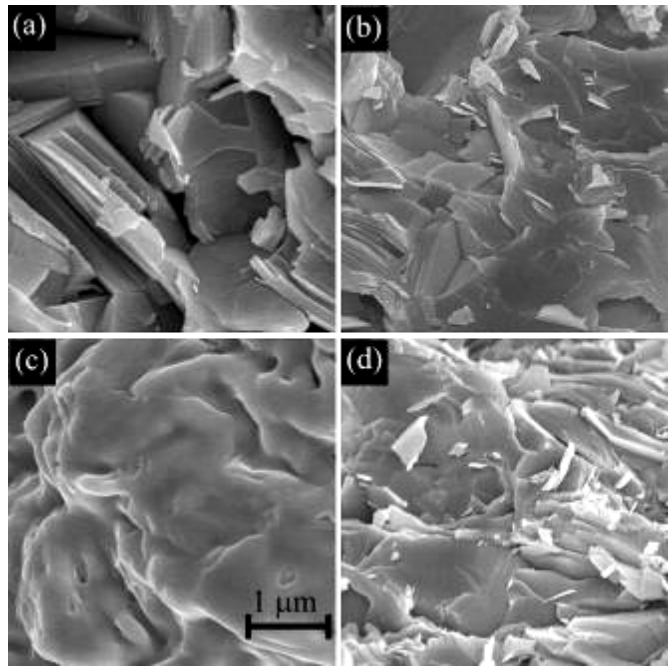

Fig. 1. SEM images of sintered BSCCO pellets: (a) starting material; (b) 2% wt slurry sonicated for 2 hours; (c) BSCCO+Pb nanocomposite; (d) BSCCO+Ag+Pg nanocomposite.

mesh, *Alfa Aesar*) in 20 mL of decane was ultrasonically irradiated for 2 hours at 263 K under a 30 mL/min flow of argon using a direct-immersion ultrasonic horn (Sonics & Materials VCX-750 at 20 kHz, ~50 W/cm$^2$). To produce composite materials, 54 μmol of powdered lead and/or silver (99.9% metals basis, 325 mesh, *Alfa Aesar*) were admixed to BSCCO slurry and sonicated under the same conditions. Resulting ultrasonically treated powders were collected by filtration, washed with dry pentane (30 mL×5), and air-dried overnight. Dry powders were pelletized at room temperature at a pressure of 2 GPa for 24 hours, maintaining an average sample mass of ~70 mg. In order to compare the results with other studies, we have used here a standard annealing procedure (850 °C in air for 48 hours, followed by a rapid quenching to the room temperature). We note, however, that properties of nanocomposites can be further optimized by modifying the annealing protocol (e.g., constant-temperature melting and recrystallization in switching atmosphere).

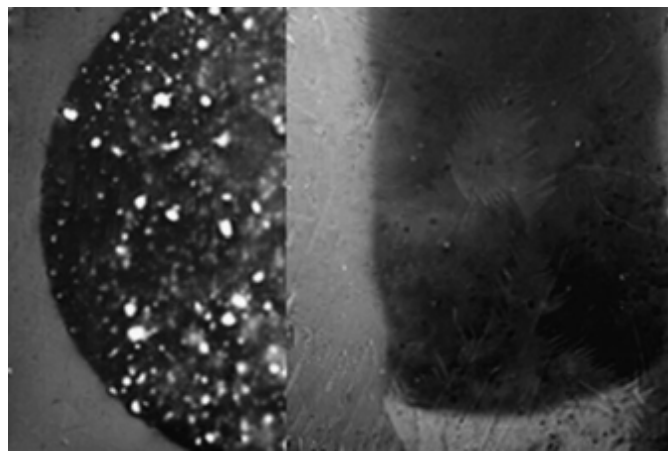

Fig. 2. Magneto-optical images of trapped flux in sintered BSCCO pellets. <u>Left:</u> initial material; <u>Right:</u> 2% wt slurry sonicated for 2 hours. (intensity is proportional to the magnitude of magnetic induction). The images are obtained as a combination of trapped flux after field-cooling and applied negative field to reveal the best contrast.

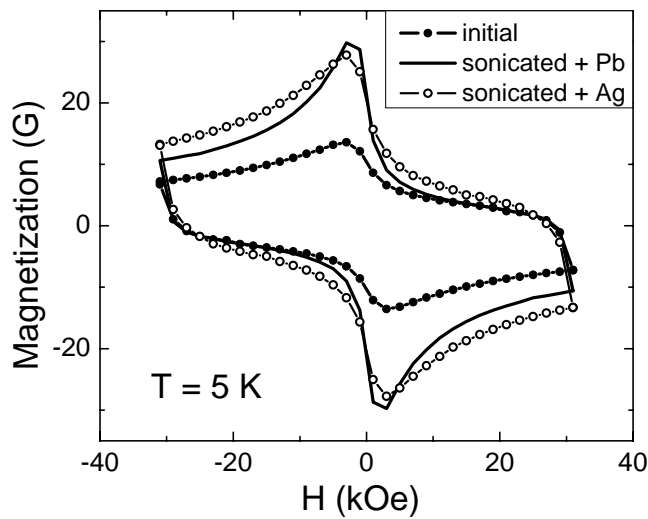

Fig. 3. Normalized magnetization loops for sintered BSCCO pellets: starting material (solid symbols), 2% wt BSCCO slurry co-sonicated for 2 hours with Pb (solid line); similar sonication with Ag (open symbols).

Slurry of 2 wt% polycrystalline YBa$_2$Cu$_3$O$_{7-\delta}$ (325 mesh, *Alfa Aesar*) in 20 mL of decane was ultrasonically irradiated for 2 hours at 263 K under a 30 mL/min flow of argon, using a direct-immersion ultrasonic horn (Sonics & Materials VCX-750 at 20 kHz, ~50 W/cm$^2$). Ultrasonically treated powders were collected by filtration, washed with dry pentane (30 mL×5), and air-dried overnight. To sustain the necessary sonochemical conditions while introducing controlled amount of oxygen into a reaction vessel, sonochemical irradiation of YBCO slurries in ethylene glycol was performed under partial oxygen flow (Ar: 20 mL/min, O$_2$: 10 mL/min). To produce composite materials, 2% YBCO slurry was sonicated with 18 μmol F(CO)$_5$ and 180 μmol of F(CO)$_5$, respectively. Resulting ultrasonically treated powders were collected by filtration, washed with dry pentane (30 mL×5), and air-dried overnight. Dry powders were pelletized at room temperature at a pressure of 2 GPa for 24 hours, maintaining an average sample mass of ~50 mg. No post-sonication annealing was performed for YBCO samples.

*C. Measurements and Characterization*

All samples were characterized by scanning electron microscope (SEM) imaging, powder x-ray diffraction, localized energy-dispersive x-ray spectroscopy (EDX), x-ray photoelectron spectroscopy (XPS), thermogravimetric analysis and differential scanning calorimetry. Morphology of superconducting powders was examined on a Hitachi S-4700 SEM equipped with an Energy Dispersive X-Ray Analysis unit. Surface chemical composition of modified powders was monitored by using X-ray Photoelectron Spectroscopy (XPS) and localized EDX. XPS analysis was conducted on a Physical

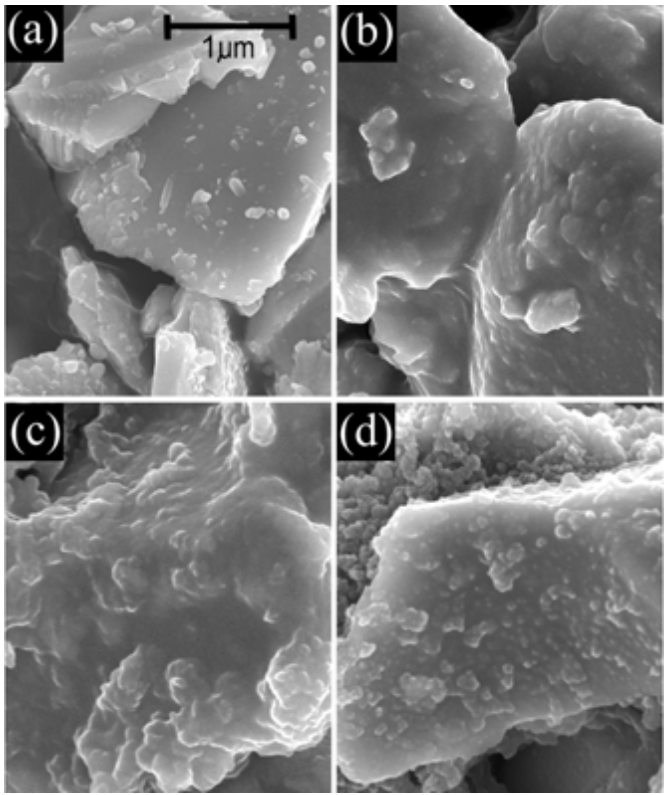

Fig. 4. Effect of ultrasound on YBCO: (a) initial material; (b) sonicated at 2% slurry load; (c) sonicated with 18 μmol of $Fe(CO)_5$; (d) sonicated with 180 μmol of $Fe(CO)_5$

Electronics PHI 5400 X-Ray Photoelectron Spectrometer, maintaining the pressure below $2.5\times10^{-8}$ torr.

Magnetic measurements were performed on *Quantum Design* MPMS. Magneto-optical imaging was done on a custom-built system with Bi-doped iron garnet in-plane Faraday indicator films. Transport measurements were performed with a standard 4-probe technique on a *Quantum Design* PPMS.

## III. RESULTS

### A. BSCCO-based Nanocomposites

SEM images of pelletized BSCCO-based nanocomposites are shown in Fig.1 – (a) BSCCO before irradiation with ultrasound; (b) BSCCO sonicated at 2% wt slurry for two hours; (c) BSCCO sonicated with lead powder; (d) BSCCO sonicated with lead and silver powders. Clearly, there is a dramatic change in sample morphology with the most prominent change for Pb-based nanocomposite. Magneto-optical measurements provide information about local distribution of the magnetic induction on the surface of superconductors [6]. Fig.2 shows magneto-optical images of the pellet made of the starting material (left) compared to that of the sonicated material (right). The grey-scale intensity in this figure is proportional to the magnetic induction. To achieve the best contrast, a combination of the flux trapping after field-cooling and application of a small negative magnetic field was used. Bright spots on the left image are the places of trapped flux, most probably inside the larger grains. The sonicated sample (right image) shows a much more uniform Meissner screening and is, therefore, more homogeneous, which correlates well with morphological changes observed in Fig.1. Since irradiated with ultrasound slurry contains both ceramic powder and soft metal granules, interparticle collisions lead to a significant size reduction and plastic deformation of softer metallic component. The latter then acts as welding or soldering material, further improving the intergrain coupling at the point of contact. Better intergrain coupling leads to the enhanced magnetic properties of sonicated BSCCO [5]. However, the effect is even more pronounced in nanocomposites made with superconducting lead and non-superconducting silver (not shown here). Magnetic measurements, shown in Fig.3, demonstrate more than two-fold enhancement of the magnetic irreversibility in nanocomposites as compared to non-modified BSCCO. The problem is that this enhancement is significant only below ~30 K and quickly diminished above. A number of factors can be responsible for this behavior, among which are non-uniform thermal expansion of nanocomposite and dimensional crossover of flux pinning in BSCCO [5]. However, current research indicates that with proper modification of the synthesis and annealing protocols, the observed enhancement can be extended to higher temperatures.

### B. YBCO-based Nanocomposites

Morphological changes induced by ultrasonic treatment in YBCO, Fig.4, are similar to $MgB_2$ [4] and BSCCO nanocomposites. However, superconducting properties did not significantly improve and even somewhat deteriorated. It became apparent that notorious sensitivity of $YBa_2Cu_3O_{7-\delta}$ to the oxygen content was the primary reason. In polycrystalline YBCO, grain boundaries are usually more oxygen deficient, compared to the bulk. Indeed, probing the sonicated materials with XPS revealed distinct changes in the chemical *surface* composition of sonicated YBCO. A single O 1s peak of the starting material, Fig.5(a), is broadened in $YBa_2Cu_3O_{7-\delta}$ irradiated with high-intensity ultrasound, and its intensity decreases. Additional lower-energy O 1s peak, Fig.5(b), can be attributed to formation of new surface layers [7,8]. Appearance and growth of oxygen peak with lowered energy, and in drop of the original oxygen peak intensity in the samples examined with XPS, confirms decrease of the surface oxygen concentration. Thus, despite the obvious rounding and fusion of the individual grains seen in SEM images of sonochemically irradiated material, its overall structure becomes chemically less homogeneous. Apparently, ultrasound irradiation apparently disrupts oxygen content in the melted surface layers, forming non-superconducting phases. In particular, formation of trace amounts of surface $Y_2BaCuO_5$ phase and $BaCuO_2$ compound within the bulk $YBa_2Cu_3O_7$ matrix has been previously reported [3]. These freshly formed non-superconducting layers enfold the superconducting grains, leading to the weakening of intergrain coupling and to reduction of the total volume of superconducting phase. However, this adverse effect can be minimized and ultimately reverted by adjusting the synthesis protocol. Sonochemical reactions are normally carried under the flow of inert [2]. Since there is no excess oxygen in the reaction vessel, oxygen content in the surface layers of superconducting $YBa_2Cu_3O_{7-\delta}$ grains inevitably decreases. To maintain the oxygen content in sonochemically treated $YBa_2Cu_3O_{7-\delta}$, irradiation of slurries was performed under partial oxygen flow in an oxygen rich solvent. It was assumed that during the sonolysis, a fraction of diatomic oxygen molecules undergoes dissociation to yield highly reactive atomic oxygen species. Remarkably, such treatment not only inhibited the oxygen loss, but apparently allowed effective saturation of surface layers with oxygen. This conclusion was ultimately confirmed by the magnetic measurements: transition temperature of YBCO sonicated under partial oxygen flow increased, compared to the initial material. Fig.5(c) shows XPS spectra for sonicated YBCO treated with ultrasound in partial oxygen flow. The undesirable low-energy peak, attributed to formation of surface non-superconducting phases, almost disappeared and the peak corresponding to the oxygen in Cu-O planes was recovered. Measurements of the magnetic moment after zero-field cooling were performed in a 10 Oe external magnetic field. The normalized by the value at 5 K, the results are shown in Fig.6. The inset shows zoomed region in the vicinity of the transition temperature clearly demonstrating the enhancement of the overall superconducting behavior.

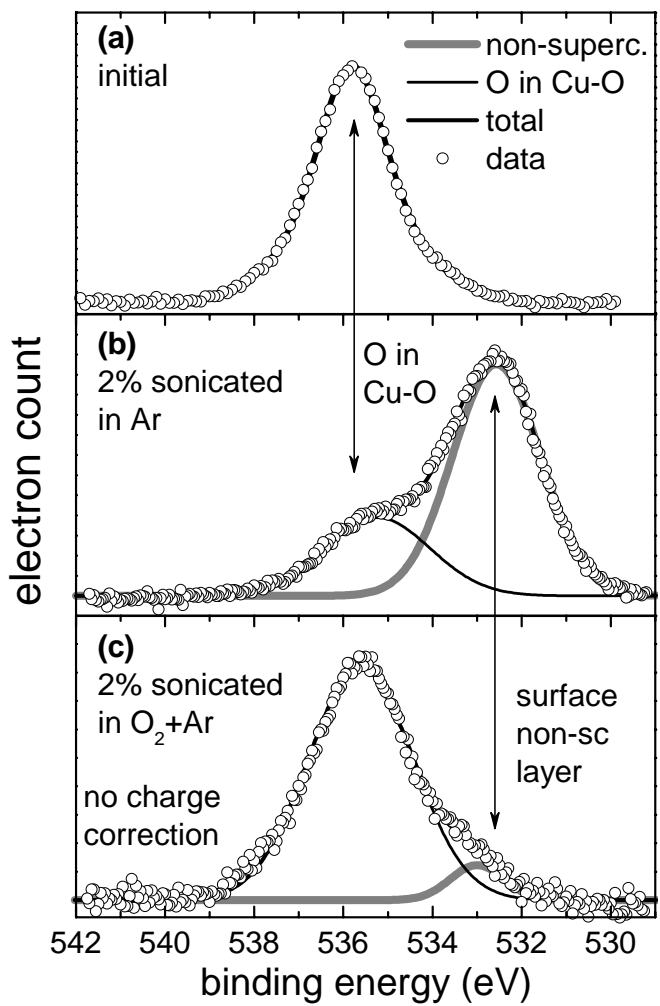

Fig. 5. Sonochemically induced enhancement of the transition temperature in YBCO: (a) starting material; (b) sonicated in Ar flow; (c) sonicated under the partial oxygen atmosphere.

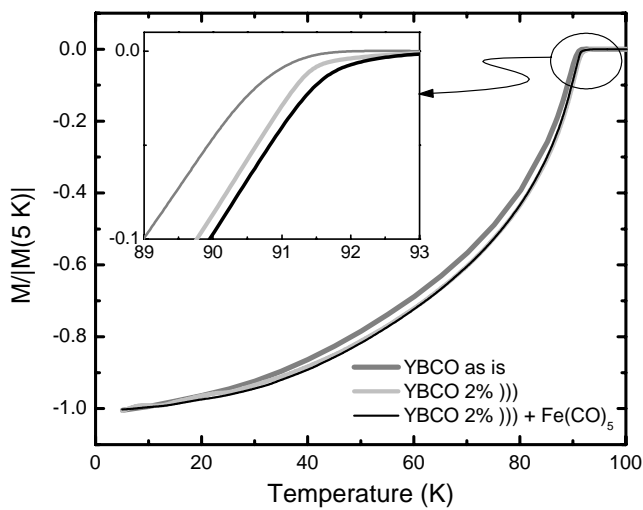

Fig. 6. Sonochemically induced enhancement of the transition temperature in YBCO measured after zero-field cooling in a 10 Oe magnetic field.

ACKNOWLEDGMENT

Discussions with K. S. Suslick, B. Ivlev, A. Gurevich and A. Polyanskii are greatly appreciated.

REFERENCES

[1] D. Larbalestier, A. Gurevich, D. M. Feldmann, and A. Polyanskii, "Superconductors: Pumping up for wire applications," Nature, vol. 414, p. 368, 2001.
[2] K. S. Suslick and G. J. Price, "Applications of Ultrasound to Materials Chemistry," J. Ann. Rev. Mat. Sci., vol. 29, p. 295, 1999.
[3] M. Gasgnier, L. Albert, J. Derouet, and L. Beaury, "Ultrasound effects on various oxides and ceramics: macro- and microscopic analyses," J. Solid State. Chem., vol. 115, p. 532, 1995.
[4] T. Prozorov, R. Prozorov, A. Snezhko, and K. S. Suslick, "Sonochemical Modification of the Superconducting Properties of MgB2," Appl. Phys. Lett., vol. 83, p. 2019, 2003.
[5] T. Prozorov, B. McCarty, Z. Cai, R. Prozorov, and K. S. Suslick, "Effects of High Intensity Ultrasound on BSCCO-2212 Superconductor," Appl. Phys. Lett., vol. 85, in print, 2004.
[6] C. Jooss, J. Albrecht, H. Kuhn, S. Leonhardt, and H. Kronmueller, "Magneto-optical studies of current distributions in high-Tc superconductors," Rep. Prog. Phys., vol. 65, p. 651, 2002.
[7] S. Evans, "Estimation of the uncertainties associated with XPS peak intensity determination," Surface and Interface Analysis, vol. 18, p. 323, 1992.
[8] A. M. Salvia and J. E. Castle, "The intrinsic asymmetry of photoelectron peaks: dependence on chemical state and role in curve fitting," Journal of Electron Spectroscopy and Related Phenomena, vol. 95, p. 45, 1998.